\documentclass[12pt,preprint]{aastex}

\shorttitle{First PPMXL photometric analysis of open cluster "Ruprecht 15"}
\shortauthors{Tadross, A. L.}

\begin{document}

\title{First PPMXL photometric analysis of open cluster "Ruprecht 15"}

\author{A. L. Tadross}
\affil{National Research Institute of Astronomy and Geophysics, Helwan, Cairo, Egypt}
\email{altadross@yahoo.com}

\begin{abstract}
We present here our first series in studying the astrophysical parameters of the open cluster "Ruprecht 15" using PPMXL\footnote{\it http://vizier.cfa.harvard.edu/viz-bin/VizieR?-source=I/317} database. In this context, the photometric, astrometry and statistical parameters for this cluster (limited radius, core and tidal radii, distances, membership, reddening, age, luminosity function, mass function, total mass, and the dynamical relaxation time) are determined for the first time.
\end{abstract}

\keywords{open clusters and associations -- individual: Ruprecht 15 -- astrometry -- Stars -- astronomical databases: catalogues.}

\section{Introduction}
Open star clusters (OCs) play an important role in studying the formation and evolution of the Galactic disk and the stellar evolution as well. Also, OCs are the most suitable objects for studying the space-age structure of the Galactic disk. The fundamental physical parameters of OCs, e.g. distance, reddening, age, and metallicity are necessary for studying the Galactic disk. The Galactic, radial and vertical, abundance gradient can be studied by OCs (Hou et al. 2000; Chen et al. 2003; Kim and Sung 2003; Tadross 2003; Kim et al. 2005). They are excellent probes of the Galactic disk structure (Janes \& Phelps, 1994; Bica et al. 2006). The strong interest of OCs results come from their fundamental properties. Among the 1787 currently OCs, most than half of them have been poorly studied or even unstudied up to now, Piatti et al. (2011). The current paper is thus part of our continuation series whose goal is to obtain the main astrophysical properties of previously unstudied OCs using modern databases. The most important thing for using PPMXL database lies in containing the positions, proper motions of USNO-B1.0\footnote{\it http://vizier.cfa.harvard.edu/viz-bin/VizieR?-source=I/284} and the Near Infrared (NIR) photometry of the Two Micron All Sky Survey (2MASS)\footnote{\it http://vizier.cfa.harvard.edu/viz-bin/VizieR?-source=II/246}, which let it be the powerful detection of the star clusters behind the hydrogen thick clouds those concentrate on the Galactic plane.

The only available information about Ruprecht 15 (hereafter Ru 15) are the coordinates and the optical apparent diameter, which were obtained from WEBDA\footnote{\it http://obswww.unige.ch/webda} site and the last updated version of DIAS\footnote{\it http://www.astro.iag.usp.br/$\sim$wilton/} collection (version 3.0, 2010 April 30). This cluster is situated in the Southern Milky Way at 2000.0 coordinates $\alpha=07^{h} \ 19^{m} \ 34^{s}, \ \delta=-19^{\circ} \ 37^{'} \ 30^{''}, \ \ell= 233.54^{\circ}, \ b= -2.896^{\circ}$, and its diameter is about 2.0 arcmin. Fig. 1 represents the blue image of Ru 15 as taken from Digitized Sky Survey (left panel), while the right panel represents J-image of the cluster as taken from Interactive 2MASS Image Service\footnote{\it http://irsa.ipac.caltech.edu/applications/2MASS/IM/interactive.html}.

This paper is organized as follows. PPMXL data extraction and preparation are presented in Section 2, while the data analysis and parameters estimations are described in Sections 3. Finally, the conclusion is devoted to Section 4.

\begin{figure}
  \centering
\includegraphics[width=0.75\textwidth]{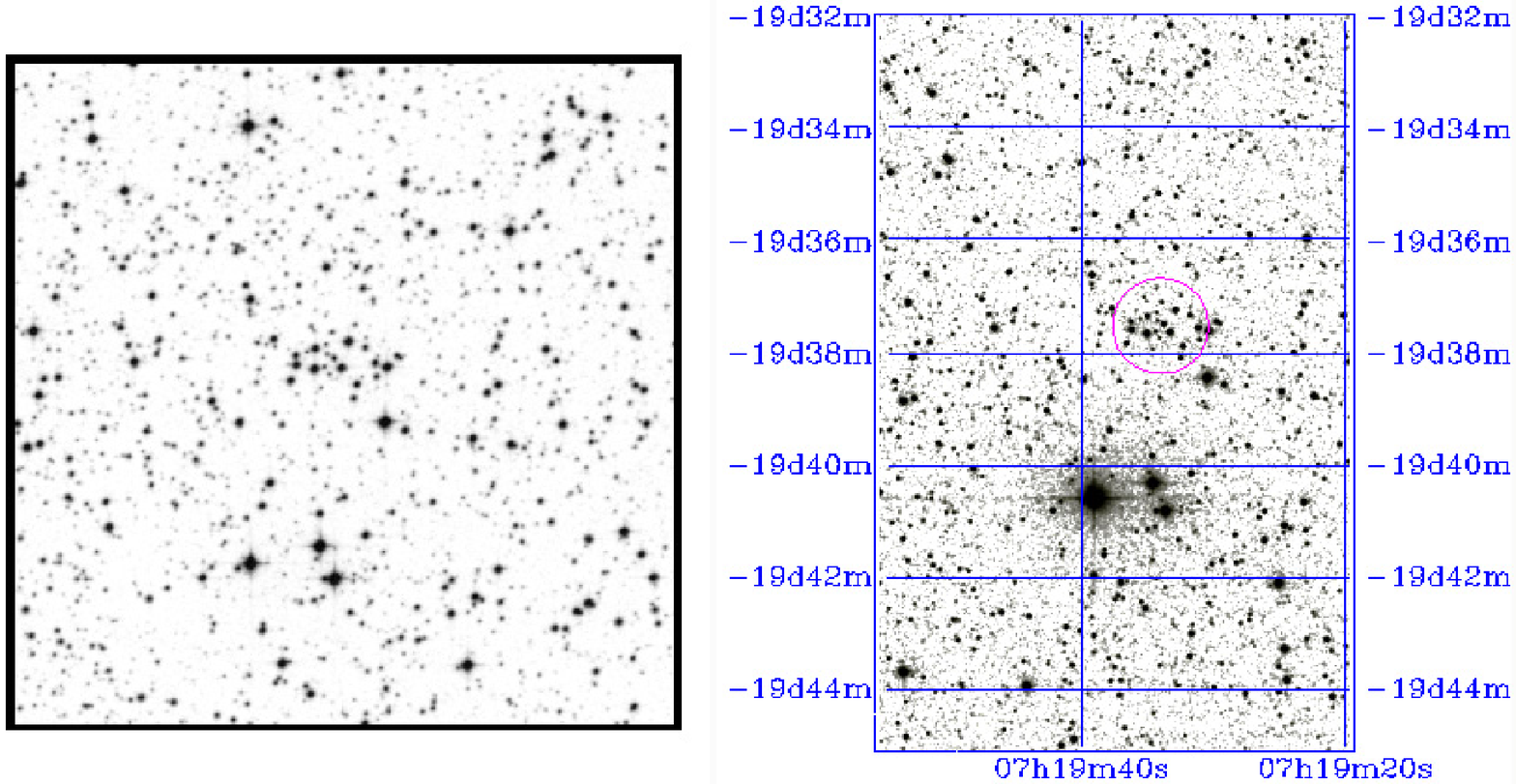}
   \caption{Left panel represents the blue image of Ru 15 as taken from Digitized Sky Surveys, while the right panel represents J-image of the cluster as taken from Interactive 2MASS Image Service, the small open circle indicates the cluster's central part. North is up, east on the left.}
   \end{figure}

\section{PPMXL Data Extraction and preparation}

The astrophysical parameters of the investigated cluster have been estimated using the PPMXL Catalogue of Roeser et al. (2010). It is combining the proper motion of USNO-B1.0 and NIR JH$K_{s}$ pass-band of 2MASS databases.

USNO-B1.0 of Monet et al. (2003) is a spatially unlimited catalogue that presents positions, proper motions, and magnitudes in various optical pass-bands. The data were obtained from scans of Schmidt plates taken for the various sky surveys during the last 50 years. USNO-B is believed to provide all-sky coverage, completeness down to V = 21 mag. It is noted that, based on the proper motion measurements, stars with large proper motions are likely to be foreground stars instead of cluster members. Background stars cannot readily be distinguished from members by proper motions. Nonetheless, identifying foreground stars is useful in cleaning up the colour-magnitude diagrams and determining the field star contamination. So, USNO-B is a very useful catalogue, which gives us an opportunity to distinguish between the members and background/field stars.

On the other hand, the NIR 2MASS photometry provides some direct answers to questions on the large scale structure of the Milky Way Galaxy and the local Universe. It provides J, H, and $K_{s}$ band photometry for millions of galaxies and nearly a half-billion stars (Carpenter, 2001). 2MASS observations are obtained using two highly automated 1.3-m telescopes, one at Mount Hopkins in Arizona (Northern Survey) and the other at Cerro Tololo in Chile (Southern Survey). Each telescope was equipped with three-channel camera, each channel consisting of 256 x 256 array HgCdTe detectors. It is uniformly scanning the entire sky in the three NIR bands J (1.25$\mu$m), H (1.65$\mu$m) and $K_{s}$ (2.17$\mu$m). This survey has proven to be a powerful tool in the analysis of the structure and stellar content of open clusters (Bica et al. 2003, Bonatto \& Bica 2003). The photometric uncertainty of the 2MASS data is less than 0.155 at $K_{s} \sim 16.5$ magnitude which is the photometric completeness for stars with $|b| > 25^{o}$, Skrutskie et al. (2006).

It is noted that Ru 15 is located near the Galactic plane ($|b|<3^{\circ}$), therefore we expect significant foreground and background field contamination. The apparent diameter is less than 5 arcmin, hence the downloaded data has been extended to reach the field background stars, whereas the cluster dissolved there, so that we extracted the data to a radius of about 10 arcmin.

Comparing the data with Cutri et al.'s Point Source Catalogue (2003), we found that most stars' magnitudes have a ``AAA'' quality flag, which means the Signal Noise Ratio is $SNR \geq 10$, i.e. they have the highest quality measurements. Every star in our data has 3-colour photometric values $J, H, K_{s}$ mag; and proper motion (pm) values in right ascension ($\alpha$) and declination ($\delta$), i.e. (pm $\alpha$ cos $\delta$ \& pm $\delta$) mas yr$^{-1}$. According to Roeser et al. (2010), the stars with proper motion uncertainties $\geq$ 4.0 mas yr$^{-1}$ have been removed.
In this context, to get a net worksheet data for Ru 15, the photometric completeness limit has been applied on the photometric pass-band 2MASS data to avoid the over-sampling at the lower parts of the cluster's CMDs (cf. Bonatto et al. 2004). The stars with observational uncertainties $\geq$ 0.20 mag have been removed. Pm vector point diagram (VPD) with distribution histogram of 2 mas yr$^{-1}$ bins for (pm $\alpha$ cos $\delta$) and (pm $\delta$) have been constructed as shown in Fig. 2. The Gaussian function fit to the central bins provides the mean pm in both directions. All data lie at that mean $\pm 1~\sigma$ (where $\sigma$ is the standard deviation of the mean) can be considered as astrometric probable members. In addition, the stellar photometric membership criteria are adopted based on the location of the stars within $\pm 0.1$ mag around the ZAMS curve in the colour magnitude diagrams (CMDs).

\begin{figure}
  \centering
\includegraphics[width=0.5\textwidth]{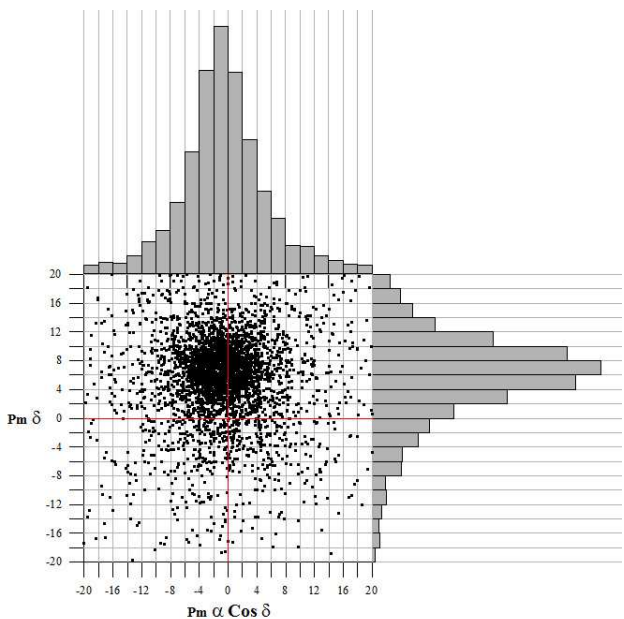}
   \caption{Proper motion vector point diagram {\it VPD} of Ru 15 after avoiding all data with pm errors $\geq$ 4 mas yr$^{-1}$. Histograms of 2 mas yr$^{-1}$ bins in both directions are drawn. The Gaussian function fit to the central bins provides the mean pm $\alpha$ cos $\delta$ = --0.84 $\pm$ 0.1 mas yr$^{-1}$ and pm $\delta$ = 6.7 $\pm$ 0.09 mas yr$^{-1}$.}
   \end{figure}

\section{Data Analysis}

\subsection{Cluster's Centre and Radial Density Profile}

The centre of any cluster can be roughly estimated by eye, but to determine the centre's coordinates of Ru 15 more precisely, we applied the star-count method to the whole area of the cluster. All the 10 arcmin data that obtained around the adopted centre are dividing into equal sized bins in $\alpha$ and $\delta$. The cluster centre is define as the location of maximum stellar density of the cluster's area. The Gaussian curve-fitting is applied to the profiles of star counts in $\alpha$ \& $\delta$ respectively as shown in Fig. 3. The differences between our estimated centre and the obtained one of Webda are shown in the figure.

To establish the radial density profile (RDP) of Ru 15, we counted the stars within concentric shells in equal incremental steps ($r\leq1$) arcmin from the cluster centre. We repeated this process for $1<r\leq2$ up to $r\leq10$ arcmin, i.e. the stellar density is derived out to the preliminary radius of the cluster. The stars of the next steps should be subtracted from the later ones, so that we obtained only the amount of the stars within the relevant shell's area, not a cumulative count. Finally, we divided the star counts in each shell to the area of that shell those stars belong to. The density uncertainties in each shell was calculated using Poisson noise statistics. Fig. 4 shows the RDP from the new centre of Ru 15 to the maximum angular separation of 5 arcmin where the stability of density has been reached. To determine the structural parameters of the cluster more precisely, we applied the empirical King model (1966). The King model parameterizes the density function $\rho(r)$ as:

\begin{equation}
\rho(r)=f_{bg}+\frac{f_{0}}{1+(r/r_{c})^{2}}
\end{equation}

where $f_{bg}$, $f_{0}$ and $r_{c}$ are background, central star density and the core radius of the cluster respectively. From the concentration parameter $c$, defined as $c= (R_{lim}/R_{core})$, Nilakshi et al. (2002) concluded that the angular size of the coronal region is about 6 times the core radius. Maciejewski \& Niedzielski (2007) reported that $R_{lim}$ may vary for individual clusters between about $2 R_{core}$ and $7 R_{core}$. In our case, we can see that $R_{lim}$=$6.9 R_{core}$, i.e. it lies within the previous vales.
The cluster limited radius can be defined at that radius which covers the entire cluster area and reaches enough stability with the background density, i.e. the difference between the observed density profile and the background one is almost equal zero. It is noted that the determination of a cluster radius is made by the spatial coverage and uniformity of PPMXL photometry which allows one to obtain reliable data on the projected distribution of stars for large extensions to the clusters' halos. On the other hand, the concentration parameter seems to be related to cluster age, i.e. for clusters younger than about 1 Gyr, it tends to increase with cluster age. Nilakshi et al. (2002) notes that the halos' sizes are smaller for older systems. Finally, we can infer that open clusters appear to be somewhat larger in the near-infrared than in the optical data, Sharma et al. (2006).
Knowing the cluster's total mass (Sec. 3.3), the tidal radius can be given by applying the equation of Jeffries et al. (2001):

\begin{equation}
R_{t} = 1.46 ~ (M_{c})^{1/3} = 10.7  \  \ pc
\end{equation}
where $R_{t}$ and $M_{c}$ are the tidal radius and total mass of the cluster respectively.

\begin{figure}
  \centering
\includegraphics[width=0.7\textwidth]{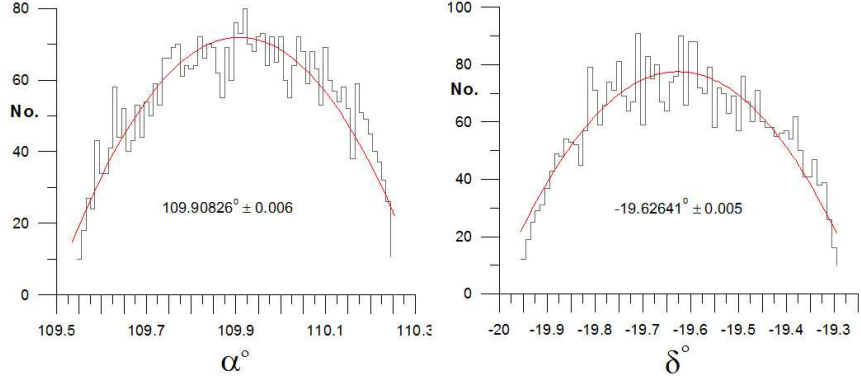}
   \caption{Estimating the cluster centre coordinates of Ru 15. The Gaussian fit provides the coordinates of highest density areas in $\alpha~ \&~ \delta$ \ as $07^{h} \ 19^{m} \ 38^{s} \ \& -19^{\circ} \ 37^{'} \ 35^{''}$ respectively. The differences between our estimated centre and the obtained one of Webda are $4^{s}$ in $\alpha$ and $5^{''}$ in $\delta$.}
   \end{figure}

%
%\clearpage
\begin{figure}
  \centering
\includegraphics[width=0.6\textwidth]{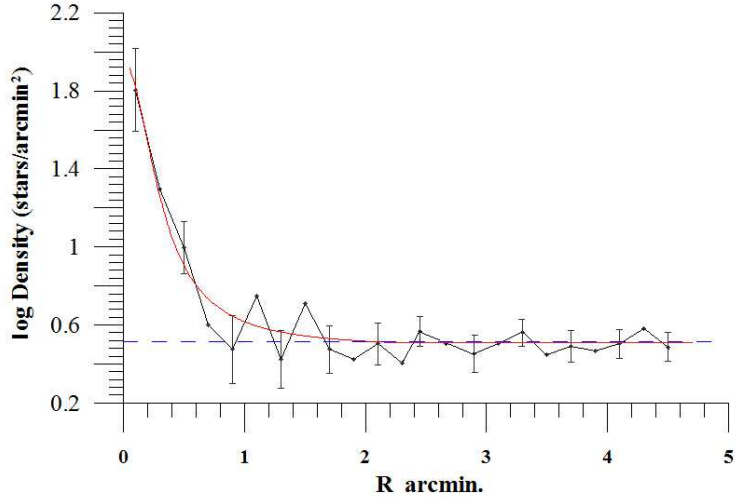}
   \caption{The radial density distribution for stars in the field of Ru 15. The density shows a maximum at the centre $\rho=63$ stars/arcmin$^{2}$ and then decreases down to $\rho=3$ stars/arcmin$^{2}$ at 2.2 arcmin, where the decrease becomes asymptotical at that point. The curved solid line represents the fitting of King (1966) model. Errors bars are determined from sampling statistics ($1/\sqrt{N}$ where N is the number of stars used in the density estimation at that point). The dashed line represents the background field density, where $f_{bg}$ = 3.0 stars per arcmin$^{2}$. The core radius $r_{c}$ = 0.32 arc min.}
   \end{figure}

%----------------------------------------------------------------

\subsection{Colour-Magnitude Diagrams}

The main photometrical parameters (age, reddening and distance) are determined for our cluster by fitting the 2MASS ZAMS solar metallicity isochrones of Marigo et al. (2008); selected and downloaded from Padova isochrones\footnote{http://stev.oapd.inaf.it/cgi-bin/cmd}; to the both CMDs (J, J-H \& K$_{s}$, J-K$_{s}$) of Ru 15. Several isochrones of different ages are applied; the best fit should be obtained at the same distance modulus for both diagrams, and the colour excesses should be obeyed Fiorucci \& Munari (2003)'s relations for normal interstellar medium as shown in Fig. 5.

The observed data has been corrected for interstellar reddening using the coefficients ratios $\frac {A_{J}}{A_{V}}=0.276$ and $\frac {A_{H}}{A_{V}}=0.176$, which were derived from absorption rations in Schlegel et al. (1998), while the ratio $\frac {A_{K_s}}{A_{V}}=0.118$ was derived from Dutra et al. (2002).

Fiorucci \& Munari (2003) calculated the colour excess values for 2MASS photometric system. We ended up with the following results: $\frac {E_{J-H}}{E_{B-V}}=0.309\pm0.130$, $\frac {E_{J-K_s}}{E_{B-V}}=0.485\pm0.150$, where R$_{V}=\frac {A_{V}}{E_{B-V}}= 3.1$. Also, we can de-reddened the distance modulus using these formulae:  $\frac {A_{J}}{E_{B-V}}$= 0.887, $\frac {A_{K_s}}{E_{B-V}}$= 0.322, then the distance of the cluster from the Sun ($R_{\odot}$) can be calculated. Therefore, (m-M)$_{J}$ =11.90 $\pm$ 0.10 mag ($\sim$ 1845 $\pm$ 85 pc), E(J-H) = 0.20 $\pm$ 0.05 mag, and (J-K$_{s}$) = 0.31 $\pm$ 0.07 mag.

Once the cluster's distance $R_{\odot}$ is estimated, then the distance from the galactic centre ($R_{g}$) and the projected distances on the galactic plane from the Sun ($X_{\odot}~\&~Y_{\odot}$) and the distance from galactic plane ($Z_{\odot}$) can be determined, see Table 1. Fig. 6 illustrates a sketch-chart of the geometric distances for a cluster in the Galaxy, obtained from Tadross (2000).

\begin{figure}
  \centering
\includegraphics[width=0.4\textwidth]{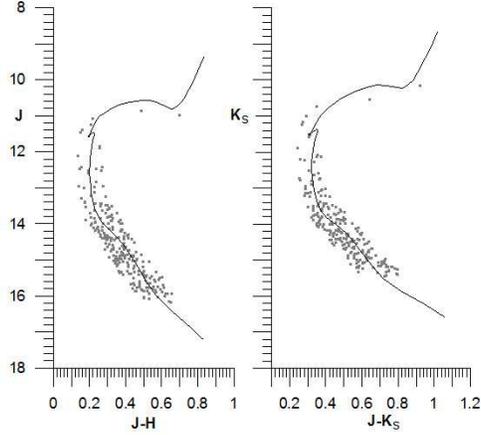}
   \caption{The net NIR CMDs of Ru 15 for stars lying closely to the fitted isochrones and after removing all contaminated field stars. Age = 500 Myr, distance modulus = 11.9 mag ($\sim$ 1845 pc), E(J-H) = 0.20 mag, and (J-K$_{s}$) = 0.31 mag.}
   \end{figure}

\begin{figure}
  \centering
\includegraphics[width=0.5\textwidth]{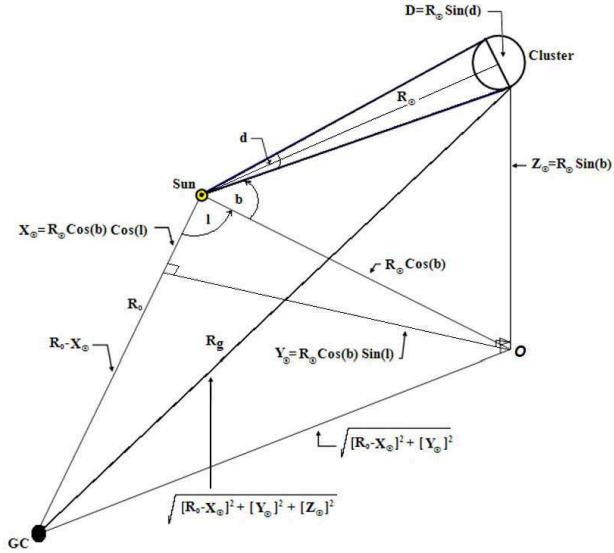}
   \caption{A sketch-chart of the geometric distances for a cluster in the Galaxy. Let (d \& D) be the angular and linear diameter in arc min and pc respectively, ($R_{\odot}$) the distance from the Sun, ($R_{\circ}$) the sun's distance from the galactic centre (8.5 kpc), ($Z_{\odot}$) the distance from the galactic plane, ($R_{g}$) the distance from the galactic centre, and ($X_{\odot}~\&~Y_{\odot}$) the horizontal projected rectangular distances of the cluster from the Sun on the galactic plane; while ($\ell~\&~b$) are the galactic longitude and latitude of the cluster.}
   \end{figure}

\subsection{Luminosity, Mass Functions and the total mass}

Since the central concentration of the cluster is relatively low, the determination of the membership using the stellar RDP is difficult. Even though we determine the membership within a circle of radius $r=2~r_{c}$ arcmin. By doing that, we obtained a more precise main-sequence in the CMDs. All of these stars are found very close to the main-sequence (MS) curve merged with some contaminated stars. These MS stars are very important in determining the luminosity and mass functions.

The luminosity function (LF) gives the number of stars per luminosity interval, or in other words, the number of stars in each magnitude bin of the cluster. It is used to study the properties of large groups or classes of objects such as the stars in clusters or the galaxies in the Local Group.
In order to estimate the LF of Ru 15 we count the observed stars in terms of absolute magnitude after applying the distance modulus as shown in Fig. 7. The magnitude bin intervals are selected to include a reasonable number of stars in each bin and for the best possible statistics of the luminosity and mass functions. From LF, we can infer that the massive bright stars seem to be centrally concentrated more than the low masses and fainter ones (Montgomery et al. 1993); the total luminosity is found to be $\sim$ $-$4.1 mag.

As known the LF and mass function (MF) are correlated to each other according to the Mass-luminosity relation. The accurate determination of both of them (LF \& MF) suffers from some problems e.g. the field contamination of the cluster members; the observed incompleteness at low-luminosity (or low-mass) stars; and mass segregation, which may affect even poorly populated, relatively young clusters (Scalo 1998). On the other hand, the properties and evolution of a star are closely related to its mass, so the determination of the initial mass function (IMF) is needed, that is an important diagnostic tool for studying large quantities of star clusters. IMF is an empirical relation that describes the mass distribution (a histogram of stellar masses) of a population of stars in terms of their theoretical initial mass (the mass they were formed with). The IMF is defined in terms of a power law as following:

\begin{equation}
\frac{dN}{dM} \propto M^{-\alpha}
\end{equation}

where $\frac{dN}{dM}$ is the number of stars of mass interval (M:M+dM), and $\alpha$ is a dimensionless exponent. The IMF for massive stars ($>$ 1 $M_{\odot}$) has been studied and well established by Salpeter (1955), where $\alpha$ = 2.35. This form of Salpeter shows that the number of stars in each mass range decreases rapidly with increasing mass. It is noted that our investigated cluster Ru 15 has MF slope ranging around Salpeter's value as shown in Fig. 8.

To estimate the total mass of Ru 15, the mass of each star has been estimated from a polynomial equation developed from the data of the solar metallicity isochrones (absolute magnitudes vs. actual masses) at the age of the cluster (500 Myr). The summation of multiplying the number of stars in each bin by the mean mass of that bin yields the total mass of the cluster, which is about 390 $M_{\odot}$.

\begin{figure}
  \centering
\includegraphics[width=0.5\textwidth]{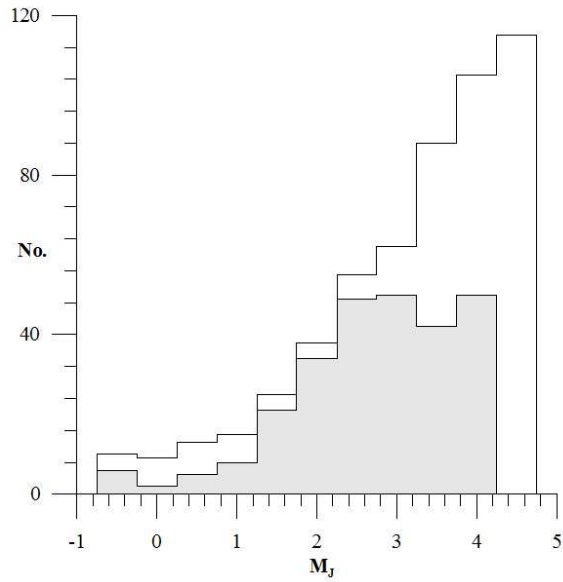}
   \caption{Spatial distribution of luminosity function of Ru 15 in terms of the absolute magnitude $M_{J}$. The colour and magnitude filters cutoffs have been applied to the cluster (dark area) and the field (white area) respectively.}
   \end{figure}

\begin{figure}
  \centering
\includegraphics[width=0.5\textwidth]{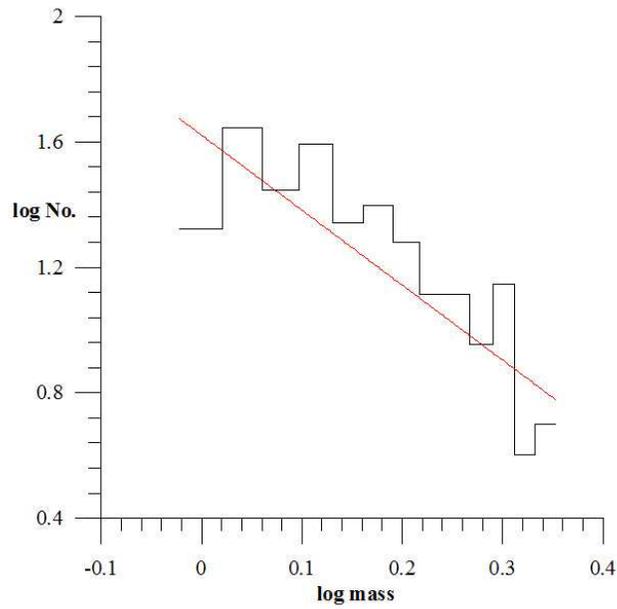}
   \caption{The mass function of Ru 15. The slope of the initial mass function {\it IMF}~ is found to be $\Gamma = -2.37 \pm 0.20$; with correlation coefficient of 0.82.}
   \end{figure}

\subsection{Dynamical state}

The relaxation time ($T_{relax}$) of a cluster is defined as the time in which the cluster needs from the very beginning to build itself and reach the stability state against the contraction and destruction forces, e.g. gas pressure, turbulence, rotation, and the magnetic field (cf. Tadross 2005).  $T_{relax}$ is depending mainly on the number of members and the cluster diameter. To describe the dynamical state of the cluster, the relaxation time can be calculated in the form:
\begin{equation}
 T_{relax}=\frac{N}{8\ln N}~T_{cross}\, , \;
\end{equation}

where $T_{cross}=D/\sigma_{V}$ denotes the crossing time, $N$ is the total number of stars in the investigated region of diameter $D$, and $\sigma_{V}$ is the velocity dispersion (Binney \& Tremaine  1998) with a typical value of 3 km s$^{-1}$ (Binney \& Merrifield 1987). Using the above formula we can estimate the dynamical relaxation time for R 15, and then the dynamical-evolution parameter $\tau$ can be calculate for the cluster by:
\begin{equation}
\tau=\frac{age}{t_{relax}}\, , \;
\end{equation}
If the cluster's age is found greater than its relaxation time, i.e.  $\tau \gg$ 1.0, then the cluster was dynamically relaxed, and vice versa. In our case, Ru 15 is indeed dynamically relaxed, where $\tau >$ 100.

\section{Conclusions}

Our procedure analysis has been applied for estimating the astrophysical parameters of yet unstudied open cluster "Ruprecht 15". Hence, we found that this cluster has real stellar density profile, its stars are lying at the same absolute distance modulus, reddening range, and having IMF slope around the Salpeter's (1955) value. On the other hand, the age of this cluster is found to be greater than the relaxation time, which infers that it is indeed dynamically relaxed. All parameters of this investigated cluster are listed in Table 1.

\begin{acknowledgements}

It is worthy to mention that, this publication made use of WEBDA, DIAS catalogues, and the data products from PPMXL database of Roeser et al. (2010).

\end{acknowledgements}

\begin{table}
%\centering
\footnotesize
\caption{The present results of Ruprecht 15.}
\begin{tabular}{ll}
\hline\noalign{\smallskip}Parameter&The present result
\\\hline\noalign{\smallskip}
Center&$\alpha$ = 07$^{h} 19^{m} 38^{s}$\\
&$\delta$ = $-$19$^{\circ} 37^{'} 35^{''}$\\
$c$ & 6.9 (see Sec. 3.1)\\
pm $\alpha$ cos $\delta$ & $-$0.84 $\pm$ 0.10 mas yr$^{-1}$\\
pm $\delta$ & ~~~~6.7 $\pm$ 0.09 mas yr$^{-1}$\\
Age& 500 $\pm$ 60 Myr.\\
Metal abundance& 0.019\\
$E(J-H)$& 0.20 $\pm$ 0.05 mag. \\
$E(J-K_{s})$& 0.31 $\pm$ 0.07 mag. \\
$E(B-V)$& 0.65 $\pm$ 0.05 mag.\\
R$_{V}$& 3.1 (see Sec. 3.2)\\
Distance Modulus& 11.90 $\pm$ 0.10 mag.\\
Distance& 1845 $\pm$ 85 pc.\\
$R_{lim}$& 2.2$^{'}$ (1.20 pc.) \\
Membership& 265 stars\\
$f_{o}$& 63 $\pm$ 2 stars/arcmin$^{2}$\\
$f_{bg}$& 3 stars/arcmin$^{2}$\\
$R_{c}$ & 0.32$^{'}\pm 0.04$ (0.20 pc)\\
$R_{t}$ & 10.7 pc.\\
$R_g$& 9.7 kpc. \\
X$_{\odot}$& 1095 kpc.\\
Y$_{\odot}$& $-$1482 kpc.\\
Z$_{\odot}$& $-$93 pc.\\
$\tau$ & $>$ 100 (see Sec. 3.4)\\
Total Luminosity & $\sim$ $-$4.1 mag.\\
{\it IMF} slope& $\Gamma$ = $-$2.37 $\pm$ 0.20\\
Total mass& $\sim$ 390 $\mathcal{M}_{\odot}$ \\
Relaxation time& 2.7 Myr\\
\hline
\end{tabular}
\end{table}

\end{document}